\newcommand{\be}{\begin{equation}}
\newcommand{\ee}{\end{equation}}
\documentstyle[12pt]{article}
\newlength{\dinwidth}
\newlength{\dinmargin}
\begin{document}

\title{
\hfill {\large IFJ 1653/PH}\\
\vspace{1 cm}
 {\bf Implications of scaling violations of
       $F_2$ at HERA for perturbative QCD}
\\[30pt]
\author{
A.J.Askew\thanks{On leave from Department of Physics,
University of Durham, England}\ ,
K.Golec\--Biernat, J.Kwieci\'nski, A.D.Martin$^*$
\\[3pt]
and
\\[3pt]
P.J.Sutton\thanks{On leave from Department of Physics,
University of Manchester, England}
\\[10pt]
{\it Department of Theoretical Physics,}
\\[3pt]
{\it Henryk Niewodniczanski Institute of Nuclear Physics,}
\\[3pt]
{\it ul. Radzikowskiego 152, 31-342 Krak\'ow, Poland }
\\[10pt]
       }
     }

\maketitle

\begin{abstract}
  We critically examine the QCD predictions for the $Q^2$
dependence of the electron-proton deep-inelastic structure
function $F_2(x,Q^2)$ in the small $x$ region, which is
being probed at HERA.  The standard results based on
next-to-leading order Altarelli-Parisi evolution are compared
with those that follow from the BFKL equation, which corresponds
to the resummation of the leading log$(1/x)$ terms.  The effects of parton
screening are also quantified.  The theoretical predictions
are confronted with each other, and with existing data from HERA.
\end{abstract}

\thispagestyle{empty}
\newpage
\setcounter{page}{1}
\vspace{1 cm}

   The first measurements of the proton structure function
$F_2(x,Q^2)$ at small $x$ have been made by the H1\cite{H1}
and ZEUS\cite{ZEUS} collaborations at HERA.  A striking increase
of $F_2$ with decreasing $x$ is observed which is consistent
with the expectations of perturbative QCD at small $x$ as
embodied in the BFKL equation\cite{BFKL}.  This equation
effectively performs a leading $\alpha_s{\rm log}(1/x)$
resummation of soft gluon emissions, which results in a
small $x$ behaviour $F_2 \sim x^{-\lambda}$ with $\lambda \sim
0.5$.

   The data at $Q^2=$ 15 and 30 GeV$^2$ are shown in Fig.1,
together with a representative set of predictions and extrapolations,
whose distinguishing features we elucidate below.  These curves fall into
two general categories.  The first, category (A), is phenomenological
and is based on parametric forms extrapolated to small $x$ with
$Q^2$ behaviour governed by the next-to-leading order Altarelli-Parisi
equations.  The parameters are determined by global fits to data
at larger $x$ (examples are the curves in Fig. 1 labelled
 MRS(D$_-^{\prime}$)\cite{MRSD}, MRS(H)\cite{MRSH} and,
to some extent, also GRV\cite{GRV}, but see below).
The second approach, denoted (B), is, in principle,
more fundamental.  Here perturbative QCD is used in the form of the
BFKL equation to evolve to small $x$ from known behaviour at larger
$x$ (e.g. AKMS\cite{AKMS}).  In other words in approach (A) the
small $x$ behaviour is input in the parametric forms used for the
parton distributions at some scale $Q^2=Q^2_0$, whereas in (B)
an $x^{-\lambda}$ behaviour at small $x$ is generated dynamically
with a determined value of $\lambda $. Of course in the phenomenological
approach, (A), it is possible to input a BFKL-motivated small $x$ behaviour
into the starting distributions (e.g. MRS(D$_-^{\prime}$) and
MRS(H) have $xg, xq_{\rm sea} \sim x^{-\lambda}$ with $\lambda
= 0.5$ and 0.3 respectively).  Since the $x^{-\lambda}$ behaviour,
for these values of $\lambda$, is stable to evolution in $Q^2$
we may anticipate that it will be difficult to distinguish approaches
(A) and (B).  However the $Q^2$ behaviour (or scaling violations) of $F_2$
is, in principle, different in the two approaches.

The Altarelli-Parisi $Q^2$ evolution is controlled by the
anomalous dimensions of the splitting functions (and by the
coefficient functions) which have been computed perturbatively
up to next-to-leading order.  On the other hand the
BFKL approach, at small $x$, corresponds to an infinite order resummation
of these quantities, keeping only leading log$(1/x)$ terms.
Summing the leading log$(1/x)$ terms, besides generating an
$x^{-\lambda}$ behaviour, gives its own characteristic $Q^2$
dependence.  One of our main purposes
is to study whether or not the
BFKL behaviour, which is more theoretically valid at small $x$,
can be distinguished from the approximate
Altarelli-Parisi parametric forms which neglect the log($1/x$) resummation.

  If we were to assume that Altarelli-Parisi evolution is valid at
small $x$ then
\begin{equation}
\frac{\partial F_2(x,Q^2)}{\partial {\rm log}Q^2} \ \simeq \ 2\sum_q e_q^2
\frac{\alpha _s(Q^2)}{2\pi}\int^1_x\frac{dy}{y}\frac{x}{y}
P_{qg}\left(\frac{x}{y} \right)
yg(y,Q^2)\ +\ ...\ ,
\end{equation}
and hence the $Q^2$ behaviour of $F_2$ can be varied by simply exploiting
the freedom in the gluon distribution at small $x$.  However the
situation is much more constrained when the BFKL equation is used to
determine the (unintegrated) gluon distribution $f(x,k^2_T)$.
Then $F_2$ may be calculated\cite{AKMS} using the $k_T$-factorization theorem
\cite{CATANI}
\begin{equation}
F_2(x,Q^2)\ =\ \int^1_x\frac{dx^{\prime}}{x^{\prime}}\int\frac{dk^2_T}
{k^4_T}f\left( \frac{x}{x^{\prime}},k^2_T \right)F^{(0)}_2(x^{\prime},
k^2_T,Q^2)
\end{equation}
where $x/x^{\prime}$ and $k_T$ are the longitudinal momentum
fraction and transverse momentum that are carried by
the gluon which dissociates into the $q\bar{q}$ pair, see Fig.\ 2.
$F^{(0)}_2$ is the quark box (and crossed box) amplitude for gluon-virtual
photon fusion\cite{AKMS}.

    In order to gain insight into the  different
possible $Q^2$ dependences of $F_2$
it is useful to introduce the moment function of the (unintegrated)
gluon distribution
\begin{equation}
f(n,k^2_T) = \int^1_0 dx x^{n-2} f(x,k^2_T).
\end{equation}
The evolution
of the moment function is given by the renormalization group
equation
\begin{equation}
f(n,k^2_T)\ =\ f(n,k^2_0)\ {\rm exp}\left[\int^{k^2_T}_{k^2_0}
\frac {dk^{\prime 2}}{k^{\prime 2}}\gamma (n,\alpha {_s}(k^{\prime2}_T))
\right]
\end{equation}
where the anomalous dimension $\gamma (n,\alpha {_s})$ is known.
{}From eq.(3)
we see that the behaviour at small $x$ is controlled by the leading
singularity of $f(n,k^2_T)$ in the $n$ plane.  In the leading
log$(1/x)$ approximation $\gamma (n,\alpha {_s})$ is just a function of
the single variable $\alpha {_s}(k^2_T)/(n-1)$ and is determined by
the BFKL kernel.  Its value is such that\cite{JAR}
\be
1\ -\ \frac{3\alpha _s(k^2_T)}{\pi (n-1)}{\tilde K}(\gamma )\ =\ 0
\ee
is satisfied, with
\be
{\tilde K}(\gamma )\ =\ 2\Psi (1)-\Psi(\gamma )-\Psi(1-\gamma ),
\ee
where $\Psi $ is the logarithmic derivative of the Euler gamma function.

For fixed $\alpha _s$  the leading singularity of $f(n,k^2_T)$
is a square root branch point at $n=1+\lambda _L$
where $\lambda _{L} =3\alpha _s{\tilde K}({1 \over 2})/\pi$
=$12\alpha _s\rm {log}2/\pi $.
Comparing with eq.(5) we find that
$\gamma (1+\lambda _{L},\alpha _s)={1\over 2}$.
Thus, from eq.(4), it directly follows that
\be
f(x,k^2_T)\sim(k^2_T)^{1\over 2} x^{-\lambda _{L}}.
\ee
Since $F^{(0)}_2/k^2_T$ in eq.(2) is simply a function of $k^2_T/Q^2$, this
leading behaviour feeds through into $F_2$ to give
\be
F_2(x,Q^2)\sim (Q^2)^{1\over 2} x^{-\lambda _{L}},
\ee
where in (7) and (8) we have omitted slowly varying logarithmic
factors.

   Formula (4) is valid for running $\alpha _s$, provided
$n$ remains to the right of the branch point throughout the
region of integration, that is provided
$n>1+12\alpha _s(k_0^2){\rm log}2/\pi$.
(For smaller values of $n$ the $k^2_T$ dependence of $f(n,k^2_T)$
is more involved\cite{JK}.)  For running $\alpha _s$ the small $x$
behaviour of $f(x,k^2_T)$ is controlled by the leading $pole$
singularity of $f(n,k_T^2)$ which occurs at $n=1+\bar{\lambda}$,
where now $\bar {\lambda}$ has to be calculated numerically\cite{KMS}.
A value of $\bar {\lambda}\approx 0.5$ is found, with rather
little sensitivity to the treatment of the infrared region of
the BFKL equation\cite{AKMS2}.
The $k^2_T$ dependence of $f$ (and hence the $Q^2$ dependence of
$F_2$) is determined by the residue $\beta$ of this pole.
Using eq.(4) we have
\be
f\sim {\beta (k^2_T)}x^{-\bar {\lambda}}
\ee
where
\be
\beta (k^2_T)\sim {\rm exp}\left[ \int^{k^2_T}_{k^2_0}
\frac {dk^{\prime 2}_T}{k^{\prime 2}_T} \gamma (1+{\bar {\lambda}},
\alpha_s(k^{\prime 2}_T))\right].
\ee
{}From the above discussion we see that this form is valid provided
$k^2_T\ge k^2_0\ge \kappa ^2(\bar {\lambda})$, where
$\kappa ^2(\bar {\lambda})$ satisfies the implicit equation
$\bar {\lambda}=12\alpha _{s}(\kappa ^2(\bar {\lambda})){\rm log}2/\pi$.
Similarly, provided that $Q^2\ge \kappa ^2(\bar {\lambda})$,
we have
\be
F_2(x,Q^2)\sim {\beta (Q^2)}x^{-\bar {\lambda}},
\ee
up to slight modifications which result from known $Q^2$ effects embedded in
$F^{(0)}_2$.  Note that for
$Q^2\stackrel{>}{\sim} \kappa ^2(\bar {\lambda})$ we should again get
an approximate $(Q^2)^{1\over 2}$ behaviour of $F_2(x,Q^2)$,
although it may (at moderately small values of $x$) be modified
by the non-leading contributions.
  Here we are also interested in $Q^2<\kappa^2(\bar{
\lambda})$ and then the form of $\beta$ is more involved\cite{JK}.

  In the leading log$(1/x)$ approximation
the anomalous dimension, $\gamma (n,\alpha _{s})$,
is a power series in $\alpha _{s}/(n-1)$.  For the BFKL approach
$\gamma (n,\alpha _{s})$ contains the sum of all these terms.
If only the first term were retained then the $Q^2$ behaviour
would correspond to Altarelli-Parisi
evolution from a singular $x^{-\bar {\lambda}}$ gluon starting
distribution with only $g\to gg$ transitions included and
with the splitting function $P_{gg}(z)$ approximated by its
singular $1/z$ term.

   It is useful to compare the $Q^2$ dependence of $F_2$ which
results from the theoretically
motivated BFKL approach, (B), with that of the
Altarelli-Parisi $Q^2$ evolution of approach (A).
For Altarelli-Parisi evolution the $Q^2$ behaviour of $F_2$ depends on the
small $x$ behaviour of the parton starting distributions.
If we assume that the starting distributions are non-singular at small $x$
(i.e.
$xg(x,Q_0^2)$ and $xq_{\rm sea}(x,Q_0^2)$ approach a constant
limit for $x\rightarrow 0$), then the leading term, which drives
both the $Q^2$ and $x$ dependence at small $x$, is of the double
logarithmic form
\be
F_2(x,Q^2) \sim {\rm exp}\left[2{\lbrace \xi(Q_0^2,Q^2){\rm log}(1/x)\rbrace
}^{1\over 2}\right],
\ee
where
\be
\xi(Q_0^2,Q^2)=\int_{Q_0^2}^{Q^2}{dq^2\over q^2}{3\alpha_s(q^2)\over \pi}.
\ee
{}From (12) we see that, as $x$ decreases, $F_2$ increases faster
than any power of log$(1/x)$ but slower than any power of $x$.

If, on the other hand,
the starting gluon and sea quark distributions are assumed to have
singular behaviour in the small $x$ limit i.e.
\be
 xg(x,Q_0^2),xq_{\rm sea}(x,Q^2_0) \sim x^{-\lambda}
\ee
with $\lambda > 0$,
then the  structure function $F_2(x,Q^2)$ behaves as
\be
F_2(x,Q^2) \sim x^{-\lambda}h(Q^2)
\ee
where the function $h(Q^2)$ is determined by the corresponding anomalous
dimensions of the moments of the (singlet) parton distributions
 at $n=1+\lambda$, as well as by the coefficient functions.

  We emphasize again that, in contrast to the BFKL
approach, for (next-to-leading
order) Altarelli-Parisi evolution the relevant quantities which
determine $h(Q^2)$ are computed from the first (two) terms in the
perturbative expansion in $\alpha _s$.  Thus terms are neglected, which
may in principle be important at small $x$, corresponding
to the infinite sum of powers of $\alpha _s/(n-1)$ in $\gamma $
(and in the coefficient function).

  Note that in both cases (i.e. eqs.(12) and (15))
Altarelli-Parisi evolution gives a slope of the structure function,
${\partial F_2(x,Q^2)/\partial {\rm log}(Q^2)}$,
which increases with decreasing $x$.
The MRS(D$_-^{\prime}$)\cite{MRSD} and MRS(H)\cite{MRSH} extrapolations
are examples of (15), with $\lambda = 0.5$ and 0.3 respectively.
On the other hand, the behaviour of $F_2$ obtained from the GRV\cite{GRV}
partons is an example of (12).  In the GRV model the partons
are generated from a valence-like input at a very low scale,
$Q^2_0=0.3{\rm GeV}^2$ (and then the valence is matched to MRS at
much higher $Q^2$). Due to the long evolution length,
$\xi (Q_0^2,Q^2)$, in reaching the $Q^2$ values corresponding to the small
$x$ HERA data the GRV prediction tends to the double
logarithmic form of (12).
The GRV model is probably best regarded as a phenomenological
way of obtaining steep distributions at a conventional input scale,
say 4GeV$^2$, since the steepness is mainly generated in the very
low $Q^2$ region where perturbative QCD is unreliable\cite{JRF}.
  Note, however,
that the steepness is specified by the evolution and is not a free
parameter.  In fact, in the region of the HERA data, the GRV
form mimics an $x^{-\lambda}$ behaviour with $\lambda \sim 0.4$,
although for smaller $x$ it is less steep.

   To summarize, we have discussed four different ways of generating
a steep $x$ behaviour of $F_2(x,Q^2)$ at small $x$, each with its own
characteristic $Q^2$ dependence: the BFKL fixed and running
$\alpha _s$ forms, (8) and (11), the Altarelli-Parisi double
leading logarithmic form with a long $Q^2$ evolution, (12),
and finally Altarelli-Parisi evolution from a steep
$x^{-\lambda}$ input, (15).  Examples of such forms are,
respectively, the fixed and running $\alpha _s$
AKMS predictions\cite{AKMS,AKMS2}, and the GRV\cite{GRV}
and MRS(H)\cite{MRSH} extrapolations.
Their $Q^2$ dependences are compared
with each other in Fig. 3
at given values of small $x$ in the HERA regime.  For reference the
MRS(D$_-^{\prime}$)\cite{MRSD} extrapolation is also shown.
The theoretical curves are calculated either from eq.(2) (where $f$
is the complete numerical solution of the BFKL equation
obtained as described in ref.\cite{AKMS2}) or from the
full next-to-leading order Altarelli-Parisi evolution.
We also show, in Fig. 3, H1\cite{H1Q} and ZEUS\cite{ZEUS}
measurements of $F_2$ made during the 1992 HERA run, corresponding
to an integrated luminosity of 25nb$^{-1}$.  Only the
statistical errors of the data are shown.
Measurements will be made with much higher luminosity, and
at smaller $x$ values, in the future.

  Several features
of this plot are noteworthy.   First, if we compare the data with
the ``$x^{-\lambda}$ dependences'' of the Altarelli-Parisi forms
of MRS(D$_-^{\prime}$), GRV and MRS(H) (which have respectively
$\lambda $ =0.5, ``$\approx $0.4'', and 0.3), then we see that
MRS(D$_-^{\prime }$) and GRV are disfavoured. So we are left with
MRS(H), which, in fact, was devised simply to reproduce\footnote
{See also the partons of the CTEQ collaboration which have
$\lambda $=0.27\cite{CTEQ2}.} the HERA data of refs.\cite
{H1,ZEUS}.

    Second, we see that the AKMS prediction (which
pre-dated the HERA data) is, like MRS(H), in good agreement
with the $x$ and $Q^2$ dependence of the data.  In principle,
it is an absolute perturbative QCD prediction of $F_2(x,Q^2)$
at small $x$ in terms of the known behaviour at larger $x$,
but, in practice, the overall normalization depends on the
treatment of the infrared region of the BFKL
equation\cite{AKMS,AKMS2}.  We can therefore normalise
the BFKL-based predictions so as to approximately describe the data at
$x=0.0027$ by adjusting a parameter which is introduced\cite{AKMS2}
in the description of the infrared region.  For the
running $\alpha _s$  AKMS calculation, this is achieved if
the infrared parameter
$k^2_a\approx 2{\rm GeV}^2$ (with $k^2_c=1{\rm GeV}^2$), in the
notation of ref.\cite{AKMS2}.  Strictly speaking, within the
genuine leading log$(1/x)$ approximation the coupling
$\alpha _s$ should be kept fixed\footnote {The use of
running $\alpha _s$ has the advantage that then the BFKL
equation reduces to the Altarelli-Parisi equation in the
double leading logarithm approximation when the transverse
momenta of the gluons become strongly ordered.}.
We therefore also solved the BFKL equation
with fixed $\alpha _s$, choosing a value $\alpha _s$=0.25 so
as to have a satisfactory normalization.  The resulting $Q^2$
dependence of $F_2(x,Q^2)$ turned out to be almost identical to that
calculated from the solution of the BFKL equation with
running $\alpha _s$.   For clarity, we therefore
have omitted the fixed $\alpha _s$ curve from Fig. 3.
Also a
background (or non-BFKL) contribution to $F_2$
has to be included in the
AKMS calculation\footnote {To be precise, we take
$F_2({\rm background})=F_2(x_0=0.1,Q^2)(x/x_0)^{-0.08}$
\cite{AKMS2}; a form which is motivated by
``soft'' Pomeron Regge behaviour\cite{DL}.  Other
reasonable choices of the background do not change our conclusions.}; this
explains why MRS(H), with $\lambda $=0.3, and AKMS, with
$\bar {\lambda} \approx 0.5$, both give equally good
descriptions of the HERA data. However, by the smallest $x$ value shown
we see that the BFKL-based AKMS predictions for $F_2$ begin to lie
significantly above those for MRS(H), due to this difference
in $\lambda $.

   A third feature of Fig. 3 is the stronger $Q^2$ dependence
of the AKMS predictions as compared with the MRS and GRV
extrapolations which are based on Altarelli-Parisi evolution.
This we had anticipated, with a growth approaching
$(Q^2)^{1\over 2}$ for BFKL as compared with the
approximately linear log$Q^2$
behaviour characteristic of Altarelli-Parisi evolution.
In reality, at the smallest $x$ value shown we find that the
AKMS growth is reduced to about $(Q^2)^{1\over 3}$, due to the fact
that $F_2({\rm background})$ is still significant.  Although
we see that the BFKL and Altarelli-Parisi $Q^2$ behaviours
are quite distinctive, to actually distinguish between them will
clearly be an experimental challenge, particularly since
 $Q^2\stackrel {<}{\sim}15{\rm GeV}^2$ is
the kinematic reach of HERA at the lowest $x$ value shown.
Recall that  the BFKL and Altarelli-Parisi equations effectively
resum the leading log$(1/x)$ and log$(Q^2)$ contributions respectively.
Thus the BFKL equation is appropriate in the small $x$ region
where
$\alpha _s{\rm log}(1/x)\sim 1$ yet
$\alpha _s{\rm log}(Q^2/Q_0^2)\ll 1$, where $Q_0^2$ is some
(sufficiently large) reference scale.  If the latter were
also $\sim 1$ then both log$(1/x)$ and log$(Q^2/Q_0^2)$
have to be treated on an equal footing\cite{GLR}, as is done, for
instance, in the unified equation proposed by Marchesini
et al.\cite{MARCH}.  For this reason we
restrict our study of small $x$ via the BFKL equation
to the region
$5\stackrel{<}{\sim}Q^2\stackrel{<}{\sim}50{\rm GeV}^2$.
As it happens, the very small $x$ HERA data lie well within this
limited $Q^2$ interval.

   So far we have neglected the effects of parton
shadowing.  If, as is conventionally expected, the gluons
are spread reasonably uniformly across the proton then
we anticipate that the effects will be small in the HERA
regime\cite{AKMS2}.  For illustration we have therefore
shown the effects of (speculative) ``hot-spot''
shadowing, corresponding to concentrations of gluons
in small hot-spots of transverse area
$\pi R^2$ inside the proton with, say, $R=2{\rm GeV}^{-1}$.
In this case, to normalise the predictions at $x$=0.0027,
we need to take the infrared parameter $k^2_a\approx 1.5{\rm GeV}^2$.
With decreasing $x$, we see from Fig. 3, that this shadowed
AKMS prediction increases more slowly than the
unshadowed one, but that it keeps the characteristic
``BFKL $Q^2$ curvature''.

     To conclude, we have performed a detailed analysis of
the $Q^2$ dependence of the structure function $F_2(x,Q^2)$
in the small $x$ region which is being probed at HERA.
We have found that the theoretically-motivated BFKL-based
predictions do indeed lead, in the HERA small $x$ regime,
to a more pronounced
curvature of $F_2(x,Q^2)$ than those based on next-to-leading
order Altarelli-Parisi evolution. The difference is illustrated
in Fig. 3 by the comparison of the AKMS curve with that for
MRS(H).  From the figure we see that data at the smallest
possible $x$ values will be the most revealing.  The
measurements shown are from the 1992 run, but data with
much higher luminosity, and at smaller $x$, will become
available in the near future.  Clearly the experimental
identification of the characteristic BFKL $Q^2$ behaviour will pose
a difficult, though hopefully not an impossible, task.

\vspace{1cm}

\noindent {\large\bf Acknowledgements}

We thank Dick Roberts and James Stirling for valuable discussions.
Two of us (JK, ADM) thank the European Community for
Fellowships and three of us (AJA, KGB, PJS) thank the Polish
KBN - British Council collaborative research programme for partial support.
This work has also been supported in part by Polish KBN grant
no. 2 0198 91 01 and by the UK Science and
Engineering Research Council.


\newpage

\noindent{\Large \bf Figure Captions}
\begin{itemize}
\item[Fig.\ 1:]
 The measurements of $F_2(x,Q^2)$ at $Q^2 = 15$ and 30 GeV$^2$ by
the H1\cite{H1} and ZEUS\cite{ZEUS} collaborations shown
 by closed and open data points respectively, with the statistical
and systematic errors added in quadrature; the H1 and ZEUS data
have a global normalization uncertainty
of $\pm $8\% and $\pm $7\% respectively. The continuous,
dotted and dashed curves respectively correspond
to the values of $F_2$ obtained from MRS(H)\cite{MRSH},
GRV\cite{GRV} and MRS(D$_-^{\prime }$)\cite{MRSD} partons.
The curves that are shown as a sequence of small squares
(triangles) correspond to the unshadowed (strong
or ``hot-spot'' shadowing) AKMS
predictions obtained by computing
$F_2=f\otimes F_2^{(0)}+F_2({\rm background})$ as in
ref.\cite{AKMS2} and as described in the text.

\item[Fig.\ 2:]
Diagrammatic display of the $k_T$-factorization formula (2),
which is symbolically of the form $F_2=f\otimes F_2^{(0)}$, where $f$
denotes the gluon ladder and $F_2^{(0)}$ the quark box (and crossed box)
amplitude.

\item[Fig.\ 3:]
The $Q^2$ dependence of $F_2(x,Q^2)$ at small $x$ (note
the shifts of scale between
the plots at the different $x$ values, which have
been introduced for clarity). The curves are as in Fig. 1.
Also shown are the measurements of the 1992 HERA
run obtained by the ZEUS
collaboration\cite{ZEUS} (open points) and, by the H1
collaboration\cite{H1Q} using their ``electron''
analysis (closed points). Only statistical errors of the data are shown.
The ZEUS points shown on the $x$=0.00098 curves are measured at an
average $x$=0.00085.  A challenge for future experiments is to distinguish
between curves like AKMS and MRS(H), both of which give a satisfactory
description of the existing data.

\end{itemize}

\end{document}